\documentclass[12pt]{article}
\usepackage{graphicx}

\newcommand{\BR}{{\cal B}}

\newcommand{\piz}{\pi^0}

\newcommand{\LK}{\mathcal{L}}

\newcommand{\psp}{\psi(2S)}
\newcommand{\jpsi}{J/\psi}
\newcommand{\EE}{e^+e^-}
\newcommand{\MM}{\mu^+\mu^-}
\newcommand{\LL}{\ell^+\ell^-}
\newcommand{\pp}{\pi^+\pi^-}

\newcommand{\kk}{K^+K^-}

\newcommand{\beq}{\begin{equation}}
\newcommand{\eeq}{\end{equation}}
\newcommand{\beqy}{\begin{eqnarray}}
\newcommand{\eeqy}{\end{eqnarray}}
\newcommand{\bitm}{\begin{itemize}}
\newcommand{\eitm}{\end{itemize}}

\newcommand{\ppjpsi}{\pi^+\pi^- J/\psi}

\newcommand{\yones}{\Upsilon(1S)}
\newcommand{\ytwos}{\Upsilon(2S)}
\newcommand{\gR}{\gamma_{\rm R}}

\newcommand{\x}{X(3872)}

\newcommand{\cdfy}{Y(4140)}

\newcommand{\etac}{\eta_c}
\newcommand{\chicz}{\chi_{c0}}
\newcommand{\chico}{\chi_{c1}}
\newcommand{\chict}{\chi_{c2}}
\newcommand{\chicJ}{\chi_{cJ}}

\newcommand{\kkpi}{K_S^0K^+\pi^- + c.c.}

\def\napoli{Graduate School of Science\\
Nagoya University, Nagoya}
\def\support{\footnote{Work 
 supported by a Grant-in-Aid for Scientific
 Research on Innovative Areas ``Elucidation of New Hadrons with a
 Variety of Flavors'' from the ministry of Education, Culture,
 Sports, Science and Technology of Japan and a Grant-in-Aid for
 for Young Scientists (B) under contract 24740158.}}

\def\Title#1{\begin{center} {\Large #1 } \end{center}}
\def\Author#1{\begin{center}{ \sc #1} \end{center}}
\def\Address#1{\begin{center}{ \it #1} \end{center}}

\newenvironment{Abstract}{\begin{quotation}  }{\end{quotation}}
\newenvironment{Presented}{\begin{quotation} \begin{center}
             PRESENTED AT\end{center}\bigskip
      \begin{center}\begin{large}}{\end{large}\end{center} \end{quotation}}
\def\Acknowledgements{\bigskip  \bigskip \begin{center} \begin{large}
             \bf ACKNOWLEDGEMENTS \end{large}\end{center}}

\def\beq{\begin{equation}}
\def\eeq#1{\label{#1}\end{equation}}
\def\eeqn{\end{equation}}

\def\beqa{\begin{eqnarray}}
\def\eeqa#1{\label{#1}\end{eqnarray}}
\def\eeqan{\end{eqnarray}}

\let\bar=\overbar

\def\Dslash{\not{\hbox{\kern-4pt $D$}}}
\def\dslash{\not{\hbox{\kern-2pt $\del$}}}

\def\BR{\mbox{\rm BR}}

\def\msb{{\bar{\ssstyle M \kern -1pt S}}}

\begin{document}
\begin{titlepage}

\vfill
\Title{Search for charmonium and charmonium-like states in $\Upsilon(1S)$
and $\Upsilon(2S)$ radiative decays}
\vfill
\Author{ C. P. Shen\support~for the Belle Collaboration}
\Address{\napoli}
\vfill
\begin{Abstract}

Using samples of 102 million $\Upsilon(1S)$ and 158 million
$\Upsilon(2S)$ event samples collected with the Belle detector,
we report on the first search for
charge-parity-even charmonium and charmonium-like states in
$\Upsilon(1S)$ and $\Upsilon(2S)$ radiative decays. No significant $\chi_{cJ}$ or
$\eta_c$ signal is observed and 90\% C.L. limits on
$\BR(\Upsilon(nS)\to \gamma \chi_{cJ})$ ($n=1,2$ and $J=1,2,3$) are obtained.
No significant signal of any charmonium-like state is
observed. The product branching fraction limits $\BR(\Upsilon(nS)\to \gamma
X(3872))$ $\BR(X(3872)\to\pi^+\pi^-\jpsi)$,
$\BR(\Upsilon(nS)\to \gamma X(3872))$ $\BR(X(3872)\to\pi^+\pi^-\pi^0
\jpsi)$, $\BR(\Upsilon(nS)\to \gamma X(3915))$
$\BR(X(3915)\to\omega \jpsi)$, and $\BR$
$(\Upsilon(1S)\to \gamma Y(4140))\BR(Y(4140)\to \phi
\jpsi)$ ($n=1,2$) are obtained at the 90\% C.L. At the same time, $\BR(\Upsilon(2S) \to \gamma
X(4350))\BR(X(4350)\to\phi\jpsi))$ is also determined at the 90\% C.L.  Furthermore,
no evidence is found for excited charmonium states below
4.8~GeV/$c^2$.

\end{Abstract}
\vfill
\begin{Presented}
The 5th International Workshop on Charm Physics\\
14-17 May 2012, Honolulu, Hawai'i 96822
\end{Presented}
\vfill
\end{titlepage}
\def\thefootnote{\fnsymbol{footnote}}
\setcounter{footnote}{0}
%

\section{Introduction}

Experimental observations near the charm threshold strongly
suggest that the spectrum of resonances with hidden charm is
remarkably more rich than suggested by the standard
quark-antiquark template and very likely includes states where the
heavy-quark $c\bar{c}$ pair is accompanied by light quarks and/or
gluons. Lots of new charmonium-like resonances ($XYZ$ particles)
in the B factories have been observed in the final states with a
charmonium and some light hadrons. They could be candidates for
usual charmonium states, however, there are also lots of strange
properties shown from these states. These may include exotic states,
such as quark-gluon hybrids, meson
molecules, and multi-quark states~\cite{review}. Many of these new
states are established in a single production mechanism or in a
single decay mode only. To better understand them, it is necessary
to search for such states in more production processes and/or
decay modes.

States with $J^{PC}=1^{--}$ can be studied via
initial state radiation (ISR) with the large $\Upsilon(4S)$ data
samples at BaBar or Belle, or via $\EE$ collisions directly at the
peak energy at, for example, BESIII. For charge-parity-even
charmonium states, radiative decays of the narrow $\Upsilon$
states below the open bottom threshold can be examined.

The production rates of the lowest lying $P$-wave spin-triplet
($\chi_{cJ}$, $J$=0, 1, or 2) and $S$-wave spin-singlet ($\eta_c$)
states in $\Upsilon(1S)$ radiative decays have been calculated in
Ref.~\cite{ktchao},  where the former is at the part per million
level, and the latter is about $5\times 10^{-5}$. The
rates in $\Upsilon(2S)$ decays are estimated to be at the same
level~\cite{ktchao}. However, there are no such calculations or
estimations for ``$XYZ$ particles" due to the limited knowledge of
their nature.

The production rates of the $P$-wave spin-triplet $\chicJ$~($J$=0,
1, 2) and $S$-wave spin-singlet $\etac$ states in $\yones$
radiative decays have been calculated by Gao {\it et al.}; the
rates in $\ytwos$ decays are estimated to be at the same
level~\cite{ktchao}. However, there are no such calculations or
estimations for ``$XYZ$ particles" due to the limited knowledge of
their nature.

Belle ever searched for charmonium and charmonium-like states in $\Upsilon(1S)$
and $\Upsilon(2S)$ radiative decays~\cite{y1spaper, y2spaper}.
The data used in those analyses include: (1).
a 5.7~fb$^{-1}$ data sample collected at the $\Upsilon(1S)$
(102 million $\Upsilon(1S)$ events) and a 1.8~fb$^{-1}$
data sample collected at $\sqrt{s}=9.43$~GeV (continuum data); (2).
a 24.7~fb$^{-1}$ data
sample collected at the $\Upsilon(2S)$ peak (158 million $\Upsilon(2S)$ events) and a 1.7~fb$^{-1}$ data
sample collected at $\sqrt{s}=9.993$~GeV (continum data).

The numbers of the $\Upsilon(1S)$ and $\Upsilon(2S)$ events are
determined by counting the hadronic events in
the data taken at the $\Upsilon$ peak after subtracting the scaled
corresponding continuum background from the data sample collected at
the corresponding energy point. The selection criteria for hadronic events are
validated with the off-resonance data by comparing the measured
$R$ value ($R=\frac{\sigma(\EE\to hadrons)}{\sigma(\EE\to \MM)}$)
with CLEO's result~\cite{cleoR}.

In this report, we gave the combined results of search for charmonium and charmonium-like states
in $\Upsilon(1S)$
and $\Upsilon(2S)$ radiative decays in order to compare easily.

\section{Belle and KEKE}

The Belle detector operating at the KEKB asymmetry-energy $\EE$ collider~\cite{KEKB}
is described in detail elsewhere~\cite{Belle}. It is
a large-solid-angle magnetic spectrometer that consists of a
silicon vertex detector (SVD), a 50-layer central drift chamber
(CDC), an array of aerogel threshold \v{C}herenkov counters (ACC),
a barrel-like arrangement of time-of-flight scintillation counters
(TOF), and an electromagnetic calorimeter comprised of CsI(Tl)
crystals (ECL) located inside a super-conducting solenoid coil
that provides a 1.5~T magnetic field. An iron flux-return located
outside of the coil is instrumented to detect $K_L^0$ mesons and
to identify muons (KLM).

\section{Search for charmonium(-like) states}

We reconstruct $\jpsi$ signals from $\EE$ or $\MM$ candidates. In
order to reduce the effect of bremsstrahlung or final-state
radiation, photons detected in the ECL within 0.05~radians of the
original $e^+$ or $e^-$ direction are included in the calculation
of the $e^+/e^-$ momentum.  In order to improve the
$\jpsi$ momentum resolution, a mass-constrained fit is
performed for $\jpsi$ signals. Different modes have
similar $\jpsi$ mass resolutions. The $\jpsi$ signal region is
defined as $|M_{\ell^+\ell^-}-m_{\jpsi}|<30~\hbox{MeV}/c^2$
($\approx 2.5\sigma$), where $m_{\jpsi}$ is the nominal mass of
$\jpsi$. The $\jpsi$ mass sidebands are defined as
$2.959~\hbox{GeV}/c^2<M_{\ell^+\ell^-}<3.019~\hbox{GeV}/c^2$ and
$3.175~\hbox{GeV}/c^2<M_{\ell^+\ell^-}<3.235~\hbox{GeV}/c^2$, and
are twice as wide as the signal region.

\begin{figure}[htbp]
\begin{center}
\includegraphics[height=7cm,angle=-90]{y1s-gchicj.epsi}\vspace{0.2cm}
\includegraphics[height=7cm,angle=-90]{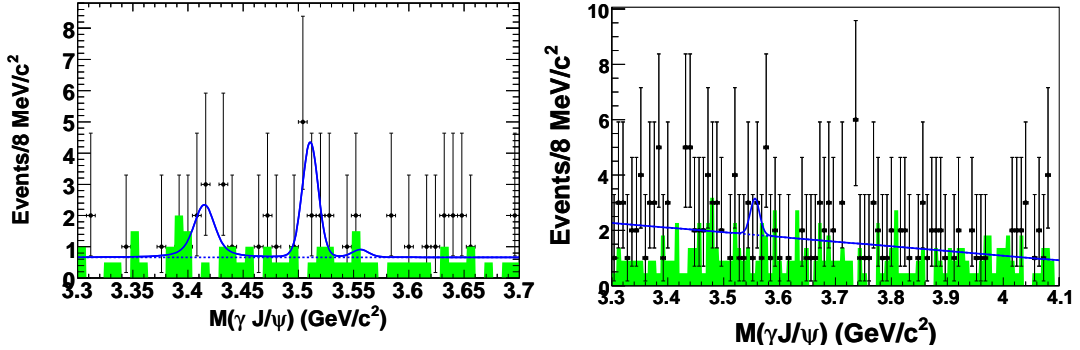}
\caption{ The $\gamma\jpsi$ invariant mass distributions
in the $\Upsilon(1S)$ (left panel) and $\Upsilon(2S)$ (right panel) data samples.
 There is no clear $\chi_{cJ}$ signal
observed. The solid curve is the best fit, the dashed curve is the
background, and the shaded histogram is from the normalized
$\jpsi$ mass sidebands.}
\label{mgll}
\end{center}
\end{figure}

We search for the $\chicJ$ in the $\gamma\jpsi$ mode. The energy
deposited by the $\chicJ$ photon is denoted as $\gamma_l$ since its
energy is much lower than that of $\gR$ (radiative photon from $\Upsilon$ decay).
The $\MM$ mode shows a clear $\jpsi$ signal, while the $\EE$ mode
has some residual radiative Bhabha background. Figure~\ref{mgll}
shows the $\gamma_l\jpsi$ invariant mass distribution together
with the background estimated from the $\jpsi$ mass sidebands
(normalized to the width of the $\jpsi$ signal range) for the
combined $\EE$ and $\MM$ modes after all the selection criteria
are applied. In Fig.~\ref{mgll}, the left panel is for the $\Upsilon(1S)$ data,
where apart from possible weak $\chi_{c0}$ and $\chi_{c1}$ signals, the
$\jpsi$ sideband events represent well the signal region, indicating
that the production of any of the $\chicJ$ states is not
significant. In Fig.~\ref{mgll}, the right panel is for the $\Upsilon(2S)$ data,
where no $\chicJ$ signal is observed. There are no structures at higher masses, where we
would expect excited $\chicJ$ states.

The upper limit on the number ($n^{\rm
up}$) of signal events at the 90\% C.L. is calculated by solving
the equation $\frac{\int_0^{n^{\rm up}}\LK(x)dx}
{\int_0^{+\infty}\LK(x)dx} = 0.9$, where $x$ is the number of
signal events, and $\LK(x)$ is the likelihood function depending
on $x$ from the fit to the data.

Bayesian upper limits on the number of events at the 90\% C.L. are found to be
$11.5$, $13.8$, and $2.4$ for the $\chicz$, $\chico$, and $\chict$ for $\Upsilon(1S)$ decay,
and $2.8$, $3.1$ and $7.6$ for the $\chicz$, $\chico$ and
$\chict$ for $\Upsilon(2S)$ decay, respectively,

\begin{figure}[htb]
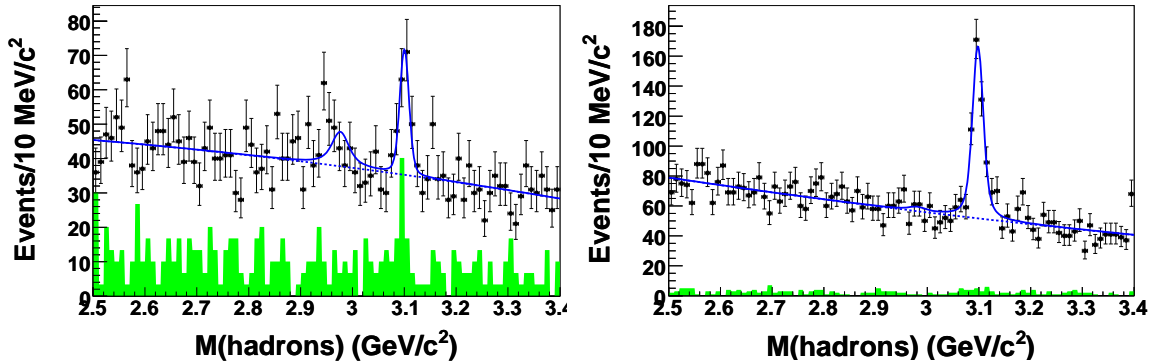

\centering
\includegraphics[height=7.5cm,angle=-90]{y1s-getac.epsi}\vspace{0.3cm}
\includegraphics[height=7.5cm,angle=-90]{y2s-getac.epsi}
\caption{ The mass distributions for a sum of the five $\eta_c$ decay modes
from $\Upsilon(1S)$ (left panel) and $\Upsilon(2S)$ (right panel) data, respectively.
The solid curve is a sum of the corresponding functions obtained
from a simultaneous fit to all the $\eta_c$ decay modes, and the
dashed curve is a sum of the background functions from the fit.
The shaded histogram is a sum of the continuum events (not
normalized in $\Upsilon(2S)$ data). The $\jpsi$ signal is produced via ISR
rather than from a radiative decay of an $\Upsilon$ resonance.}
\label{metac}
\end{figure}

To study the $\gamma\etac$ mode, we reconstruct the $\etac$ mass
from the invariant masses of $\kkpi$, $\pp\kk$, $2(\kk)$, $2(\pp)$,
and $3(\pp)$. Figure~\ref{metac} shows the combined mass distribution for the five
$\etac$ decay modes after all of the selection. The left panel is for
$\Upsilon(1S)$ data, while the right is for $\Upsilon(2S)$ data.
The peak in hadronic mass at the $\jpsi$ mass, as seen in Fig.~\ref{metac}, can
be attributed to the ISR process, $\EE \to \gamma_{{\rm ISR}}
\jpsi$, while the accumulation of events within the $\etac$ mass
region is small. The shaded histogram in Fig.~\ref{metac} is the
same distribution for the continuum data (not
normalized in $\Upsilon(2S)$ data).

A simultaneous fit is performed to the five final states.
The ratios of the $\etac$ ($\jpsi$)
yields in all the channels are fixed to $\BR_i\epsilon_i$, where each
$\BR_i$ is the $\etac$ ($\jpsi$) decay branching fraction
for the $i$-th mode reported by the PDG~\cite{PDG}, and $\epsilon_i$
is the MC-determined efficiency for this mode. The fit function
contains a BW function convolved with a Gaussian resolution
function (its resolution is fixed to 7.9~$\hbox{MeV}/c^2$ from MC
simulation) describing the $\etac$ signal shape, another Gaussian
function describing the $\jpsi$ signal shape, and a second-order
polynomial describing the background shape. The mass and width of
the BW function are fixed to the PDG values~\cite{PDG} for the
$\etac$. The fitted results are shown in Fig.~\ref{metac},
where the solid line is the sum of the best fit functions in the
simultaneous fit, and the dashed line is the sum of the background
functions.

The fits yield $46\pm 22$ and $14\pm 20$ $\Upsilon(1S) \to \gamma \etac$
and $14\pm 20$ $\Upsilon(2S) \to \gamma \etac$ signal events, respectively.
The corresponding upper limits on the
number of the $\etac$ signal events are estimated to be 72
and 44, respectively, at the
90\% C.L.

\begin{figure}[htbp]
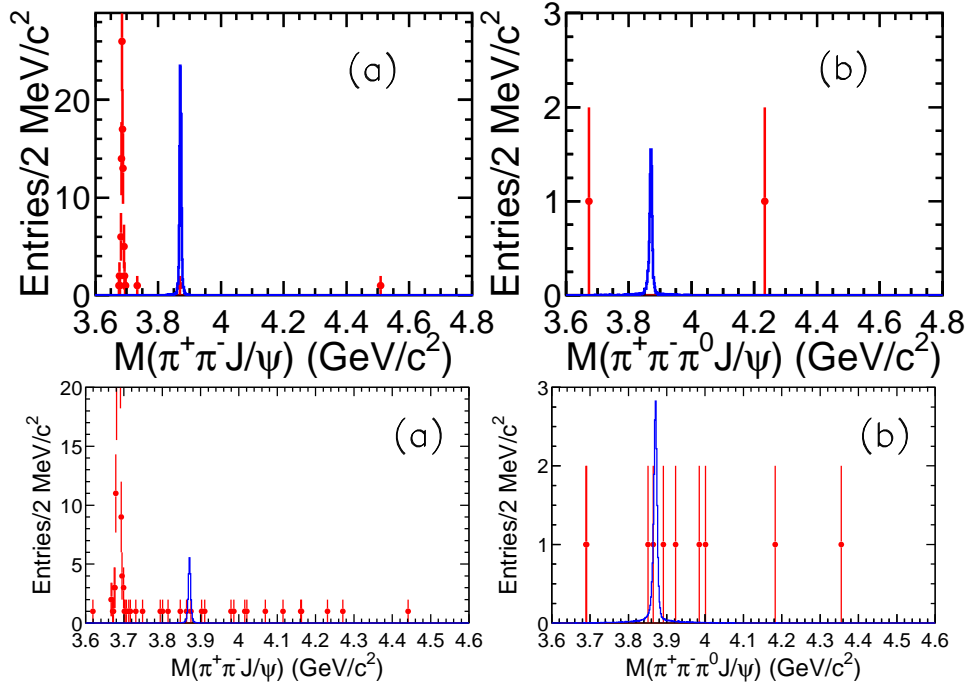

\centering
\includegraphics[height=5cm,angle=0]{y1s-x3872.epsi}
\includegraphics[height=4.cm,angle=0]{y2s-x3872-1.epsi}
\includegraphics[height=4.cm,angle=0]{y2s-x3872-2.epsi}
\caption{(a) Distributions of the $\pp \jpsi$ invariant mass. (b) Distributions of the $\pp\pi^0\jpsi$
invariant mass.
The upper two plots are for $\Upsilon(1S)$ decay, while the lower for
$\Upsilon(2S)$ decay. Points
with error bars are data, open histograms are the MC expectation for
the $X(3872)$ signal (not normalized). The peak at $3.686$~GeV/$c^2$
in (a) is due to $\psi(2S)$ production via ISR.} \label{x3872}
\end{figure}

The selection criteria for $\yones/\Upsilon(2S)\to \gamma\x$, $\x\to \ppjpsi$ are
similar to those used for ISR $\ppjpsi$ events in $\Upsilon(4S)$
data~\cite{belley}.  Except for a few residual ISR produced $\psp$ signal
events, only a small number of events appear above the $\psp$ peak
in the $\pp\jpsi$ invariant mass distribution, as shown in
Fig.~\ref{x3872} (left two plots), where the upper plot is for the
$\Upsilon(1S)$ data, while the lower is for the $\Upsilon(2S)$ data.
For $\Upsilon(1S)$, within the $X(3872)$ signal region, there is only
one event with a mass of 3.870 GeV/$c^2$.  The upper limits at 90\% C.L. on the number
of the $X(3872)$ signal events are estimated to be 3.9 and 3.6 from
$\Upsilon(1S)$ and $\Upsilon(2S)$ decays, respectively.

We validate our analysis by measuring the $\psp$ ISR production
cross section as observed in the $\ppjpsi$ mode. The measured cross section of
$\EE\to \gamma_{\rm ISR}\psp$ is in agreement with a
theoretical calculation of $18.5$~pb using PDG~\cite{PDG} values for
the $\psp$ resonance parameters as input.

We also search for the $\x$ and $X(3915)$ in the $\pp\piz\jpsi$
mode. We select $\pi^+$, $\pi^-$, and $\jpsi$ candidates in the
$\x\to\ppjpsi$ mode and a $\pi^0$ candidate
from a pair of photons with invariant mass within 10~MeV$/c^2$ of
the $\piz$ nominal mass. Here the $\piz$ mass resolution is about
4~MeV/$c^2$ from MC simulation. Figure~\ref{x3872} (right two plots)
shows the $\pp\piz\jpsi$ invariant mass distributions, where the open
histogram is the MC expectation for the $X(3872)$ signal plotted
with an arbitrary normalization. We
observe two events in the $\pp \pi^0 \jpsi$ mass spectrum between
3.6~GeV/$c^2$ and 4.8~GeV/$c^2$ in the $\yones$ data. For these two
events, the $\pp \pi^0 \jpsi$ masses are 3.67~GeV/$c^2$ and
4.23~GeV/$c^2$, and the corresponding $\pp \pi^0$ masses are
0.54~GeV/$c^2$ and 1.04~GeV/$c^2$, respectively. The event at
3.67~GeV/$c^2$, is likely to be from $\EE \to \gamma_{\rm ISR} \eta
\jpsi \to \gamma_{\rm ISR} \pp \pi^0 \LL$, since 0.9 events are
expected from MC simulation. In the $\Upsilon(2S)$ data, there are a few
events scattering in the $\pp \pi^0 \jpsi$ mass spectrum. No event is observed within the
$X(3872)$ or $X(3915)$ mass region in the $\Upsilon(1S)$ data, while
there is one event with $m(\pp \piz
\jpsi)$ at 3.923~GeV/$c^2$ and $m(\pp \piz)$ at 0.790~GeV/$c^2$
from the $\Upsilon(2S)$ data. We determine $n^{\rm up}$ for the number
of $\x$ signal events to be 2.3 and 4.2 at the 90\% C.L. for $\Upsilon(1S)$
and $\Upsilon(2S)$ decays, respectively. Assuming that
the number of background events is zero, the upper limits $n^{\rm
up}$ for the number of $X(3915)$ signal events are 2.3 and 4.4 at the 90\%
C.L. for $\Upsilon(1S)$ and $\Upsilon(2S)$ decays, respectively.

We search for the $Y(4140)$ and the $X(4350)$ (only in the $\Upsilon(2S)$ data)
in the $\phi\jpsi$ mode. The selection criteria are very similar to those in the
analysis of $X(3872)\to \ppjpsi$ and the $\phi$ is
reconstructed from a $\kk$ pair. According to MC simulation, the
$\phi$ signal region is defined as
$1.01~\hbox{GeV}/c^2<M_{\kk}<1.03$~GeV/$c^2$.  After
applying all of the above event selection criteria, there is no
clear $\jpsi$ or $\phi$ signal in $\Upsilon(1S)$ or $\Upsilon(2S)$ data samples.
Nor are there candidate events in
the $Y(4140)$ or $X(4350)$ mass regions.  The upper limits on the
number of $Y(4140)$ and $X(4350)$ signal events are both 2.3 at
the 90\% C.L.

\section{Summary}

To summarize, we find no significant signals for the $\chicJ$ or
$\eta_c$, as well as for the $\x$, $X(3915)$, $\cdfy$, or
$X(4350)$ in $\Upsilon(1S)$ and $\ytwos$ radiative decays.
In addition, we find no evidence for
excited charmonium states in the invariant mass distributions of all final states
below 4.8~GeV/$c^2$. Table~\ref{summary} lists the final results for the
upper limits on the branching fractions of all the states studied.
In order to calculate
conservative upper limits on these branching fractions, the
efficiencies have been lowered by a factor of $1-\sigma_{\rm sys}$,
here $\sigma_{\rm sys}$ is the total systematic error.
The results obtained on the $\chicJ$
and $\etac$ production rates are not in contradiction with the
calculations in Ref.~\cite{ktchao}. No $\x$, $X(3915)$, or $\cdfy$
signals are observed, and the production rates of the $\ppjpsi$,
$\pp\piz\jpsi$, $\omega \jpsi$, or $\phi\jpsi$ modes are found to be
less than a few times $10^{-6}$ at the 90\% C.L.

\begin{table}[htbp]
\caption{Summary of the limits on $\Upsilon(1S)$ and  $\Upsilon(2S)$ radiative decays to
charmonium and charmonium-like states $R$. Here $\BR(\Upsilon \to \gamma R)^{\rm up}$ (${\cal
B}_R$) is the upper limit at the 90\% C.L. on the decay branching
fraction in the charmonium state case, and on the product
branching fraction in the case of a charmonium-like state.}
\label{summary}
\begin{center}
\begin{tabular}{c  c  c }
\hline
 State ($R$)& ${\cal B}_R$ ($\Upsilon(1S)$) & ${\cal B}_R$ ($\Upsilon(2S)$) \\
\hline
 $\chi_{c0}$                   & $6.5\times 10^{-4}$ & $1.0\times 10^{-4}$ \\
 $\chi_{c1}$                   &   $2.3\times 10^{-5}$ & $3.6\times 10^{-6}$ \\
 $\chi_{c2}$                    & $7.6\times 10^{-6}$ & $1.5\times 10^{-5}$ \\
 $\eta_c$                      & $5.7\times 10^{-5}$ & $2.7\times 10^{-5}$\\
 $X(3872) \to \pp \jpsi$  & $1.6\times 10^{-6}$ &  $0.8\times 10^{-6}$\\
 $X(3872) \to \pp \pi^0 \jpsi$  & $2.8\times 10^{-6}$ & $2.4\times 10^{-6}$ \\
 $X(3915) \to \omega \jpsi$  & $3.0\times 10^{-6}$ & $2.8\times 10^{-6}$  \\
 $Y(4140) \to \phi \jpsi$ & $2.2\times 10^{-6}$ &$1.2\times 10^{-6}$ \\
 $X(4350) \to \phi \jpsi$ & $\cdots$ & $1.3\times 10^{-6}$ \\
 \hline
\end{tabular}
\end{center}
\end{table}


\Acknowledgements
We combined many material from  Ref.~\cite{y1spaper} and Ref.~\cite{y2spaper}, where the
research was done for the $\Upsilon(1S)$ and $\Upsilon(2S)$ data, respectively.
We thank the organizers for their kind invitation and
congratulate them for a successful workshop.

\end{document}